\documentclass[graybox]{svmult}

\usepackage{mathptmx}       

\usepackage{helvet}         
\usepackage{courier}        
\usepackage{type1cm}        
\usepackage{makeidx}         
\usepackage{graphicx}        
\usepackage{multicol}        
\usepackage[bottom]{footmisc}

\usepackage{cite}

\usepackage{amssymb,amsfonts,amsmath}

\makeindex             

\begin{document}

\newcommand{\EQ}{Eq.~}
\newcommand{\EQS}{Eqs.~}
\newcommand{\FIG}{Fig.~}
\newcommand{\FIGS}{Figs.~}
\newcommand{\SEC}{Sec.~}
\newcommand{\SECS}{Secs.~}

\title*{Self-exciting point process modeling of conversation event sequences}
\titlerunning{Self-exciting point process modeling}
\author{Naoki Masuda and Taro Takaguchi and Nobuo Sato and Kazuo Yano}
\institute{Naoki Masuda \at Department of Mathematical Informatics, The University of Tokyo, 7-3-1 Hongo, Bunkyo, Tokyo 113-8656, Japan, \email{masuda@mist.i.u-tokyo.ac.jp}
\and Taro Takaguchi \at Department of Mathematical Informatics, The University of Tokyo, 7-3-1 Hongo, Bunkyo, Tokyo 113-8656, Japan, \email{taro\_takaguchi@mist.i.u-tokyo.ac.jp}
\and Nobuo Sato \at Central Research Laboratory, Hitachi, Ltd., 1-280 Higashi-Koigakubo, Kokubunji-shi, Tokyo 185-8601, Japan, \email{nobuo.sato.jn@hitachi.com}
\and Kazuo Yano \at Central Research Laboratory, Hitachi, Ltd., 1-280 Higashi-Koigakubo, Kokubunji-shi, Tokyo 185-8601, Japan, \email{kazuo.yano.bb@hitachi.com}}

\maketitle

%


\section*{Abstract}
Self-exciting processes of Hawkes type have been used to model various phenomena including earthquakes, neural activities, and views of online videos. Studies of temporal networks have revealed that sequences of social interevent times for individuals are highly bursty. We examine some basic properties of event sequences generated by the Hawkes self-exciting process to show that it generates bursty interevent times for a wide parameter range. Then, we fit the model to the data of conversation sequences recorded in company offices in Japan. In this way, we can estimate relative magnitudes of the self excitement, its temporal decay, and the base event rate independent of the self excitation. These variables highly depend on individuals. We also point out that the Hawkes model has an important limitation that the correlation in the interevent times and the burstiness cannot be independently modulated.

\section{Introduction}\label{sec:introduction}

\subsection{Temporal Networks}

Social networks, which specify the pairs of individuals that are
directly connected and those that are not, are
substrates of social interactions.
An important caveat in the use of social networks for understanding social behavior is that the pair of directly connected individuals does not interact all the time. Social events between a pair of individuals, such as dialogues and
transmission of email, are 
better described as a sequence of events, i.e., a collection
of tagged event times, where the tag includes, for example,
the identity of the
two individuals, type of the event, duration, and
content of dialogues.
In fact, recent massive data, mostly online, and 
technological developments of recording devices for offline social interaction
enable recording of
social events with a higher temporal (and spatial) precision than
before. Examples of data taken in this domain include
calling activity \cite{Candia2008JPA},
web recommendation writing \cite{Iribarren2009PRL},
email traffic \cite{Eckmann2004PNAS,Barabasi2005Nature,Malmgren2008PNAS},
online forum dealing with sexual escorts \cite{Rocha2010PNAS}, human interactions in the real space
\cite{Cattuto2010PlosOne,Isella2011JTB,Isella2011PlosOne,Takaguchi2011PRX},
to name a few. Transmission of infection or information may occur
only during the period in which two individuals are involved in an event.
A set of such event sequences among pairs of individuals
are collectively called the
temporal network \cite{HolmeSaramaki2012PhysRep}, which is the focus of
this volume.
Computational models that generate realistic event sequences
possessing properties such as those described in
\SECS\ref{sub:long tail IET} and \ref{sub:IET correlation}
would help us understand
the nature of human communication behavior.

\subsection{Long-tailed Interevent Time Distribution}\label{sub:long tail IET}

\begin{figure}
\begin{center}
\includegraphics[width=6cm]{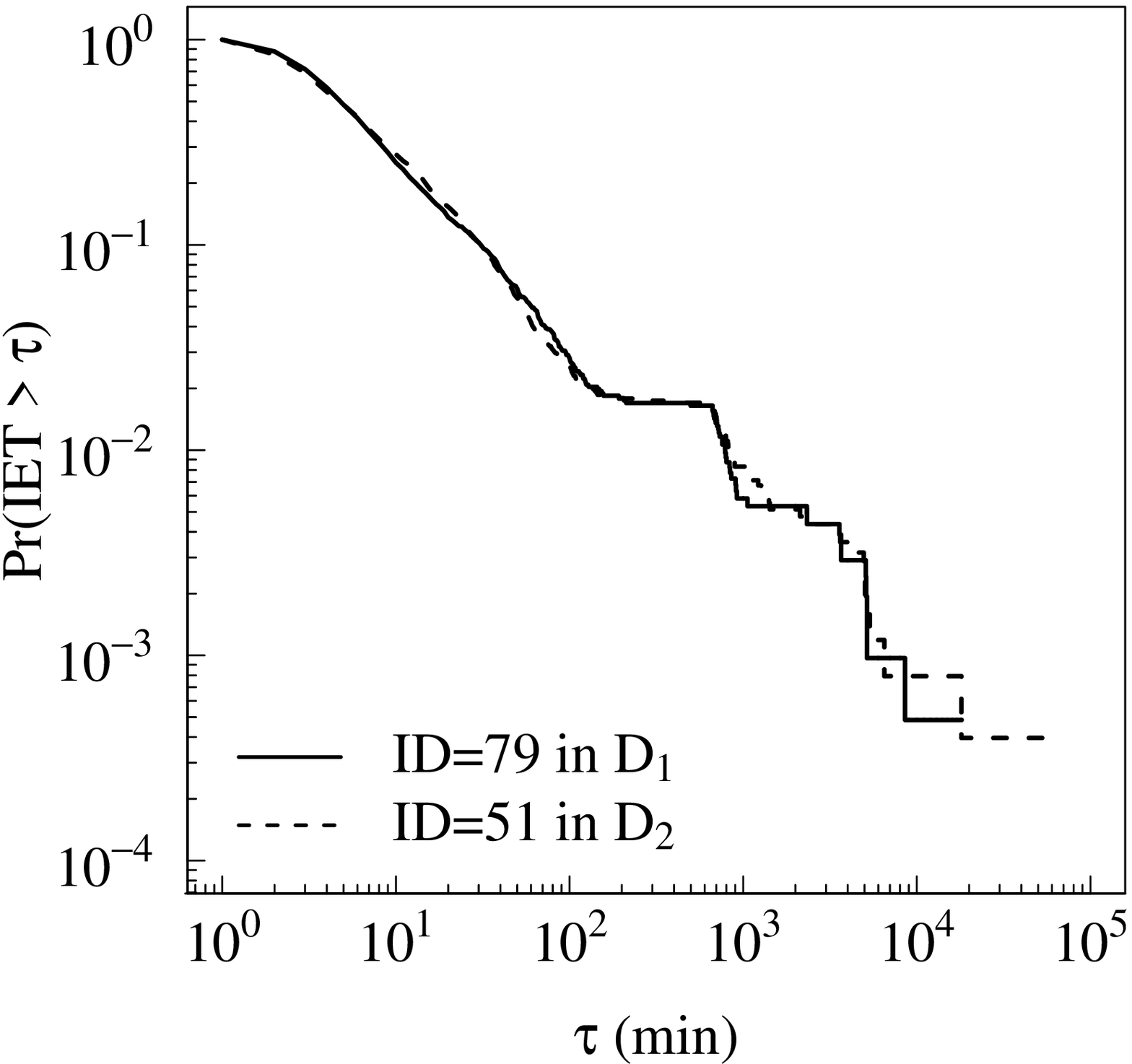}
\caption{Survivor functions of
the IET (i.e., probability that the IET is larger than $\tau$)
for the conversation sequences of two individuals.
For each of $D_1$ and $D_2$, the individual with the largest number of events
is selected. The selected individuals in $D_1$ and $D_2$ have 2,397 and 2,886 events, respectively.}
\label{fig:IET dist}
\end{center}
\end{figure}

In many empirical event sequences that we would like to model,
 interevent times are
distributed according to a
long-tailed distribution. 
The survivor functions (also called the complementary cumulative distributions)
of IET (i.e., the probability that the IET is larger than a given value $\tau$),
are shown in \FIG\ref{fig:IET dist} for two individuals in
the data sets used in our previous study \cite{Takaguchi2011PRX,Takaguchi2012NJP} (see \SEC\ref{sub:data sets} for descriptions of the data).

Different mechanisms seem to explain the non-Poissonian behavior of
the IET.
A first mechanism that
was discovered to generate power-law IET distributions is
a priority queue model \cite{Barabasi2005Nature}. In this class of models,
each task corresponding to an event carries a priority level and arrives at a queue.
Then, the queue tends to execute tasks with high priority; tasks with low priority are made to wait for a long time before being executed.
The priority queue model has also been extended to allow for
interaction of two priority queues between a pair of interacting individuals 
\cite{Oliveira2009PhysicaA,Min2009PRE,WuZhou2010PNAS,Jo2012NewJPhys}.
However, some types of social interaction including
conversations may not proceed like a queue.
Therefore, we attempt an alternative approach in the present chapter.

\subsection{IET Correlation}\label{sub:IET correlation}

Another facet of actual event sequences is that they
often possess positive temporal correlation. In other words,
a long (short) IET is likely to be followed by a long (short) IET.
This is the case even if the effect of circadian fluctuations is removed
from data \cite{Karsai2012SciRep}. Although there are various methods to measure
temporal correlation of the IET \cite{Goh2008EPL,Karsai2012SciRep}, here we show it
by simply measuring the conditional mean IET defined by
\begin{equation}
\tau^{\rm next}(\tau)\equiv \left<\tau_{i+1}\right>_{\tau_i\le \tau},
\label{eq:tau next}
\end{equation}
where $\tau_i$ is the $i$th IET in a sequence, and $\left<\cdot\right>$ represents the average. If the IET correlation is absent, $\tau^{\rm next}(\tau)$ is independent of $\tau$.

\begin{figure}
\begin{center}
\hspace*{-1cm}
\includegraphics[width=6cm]{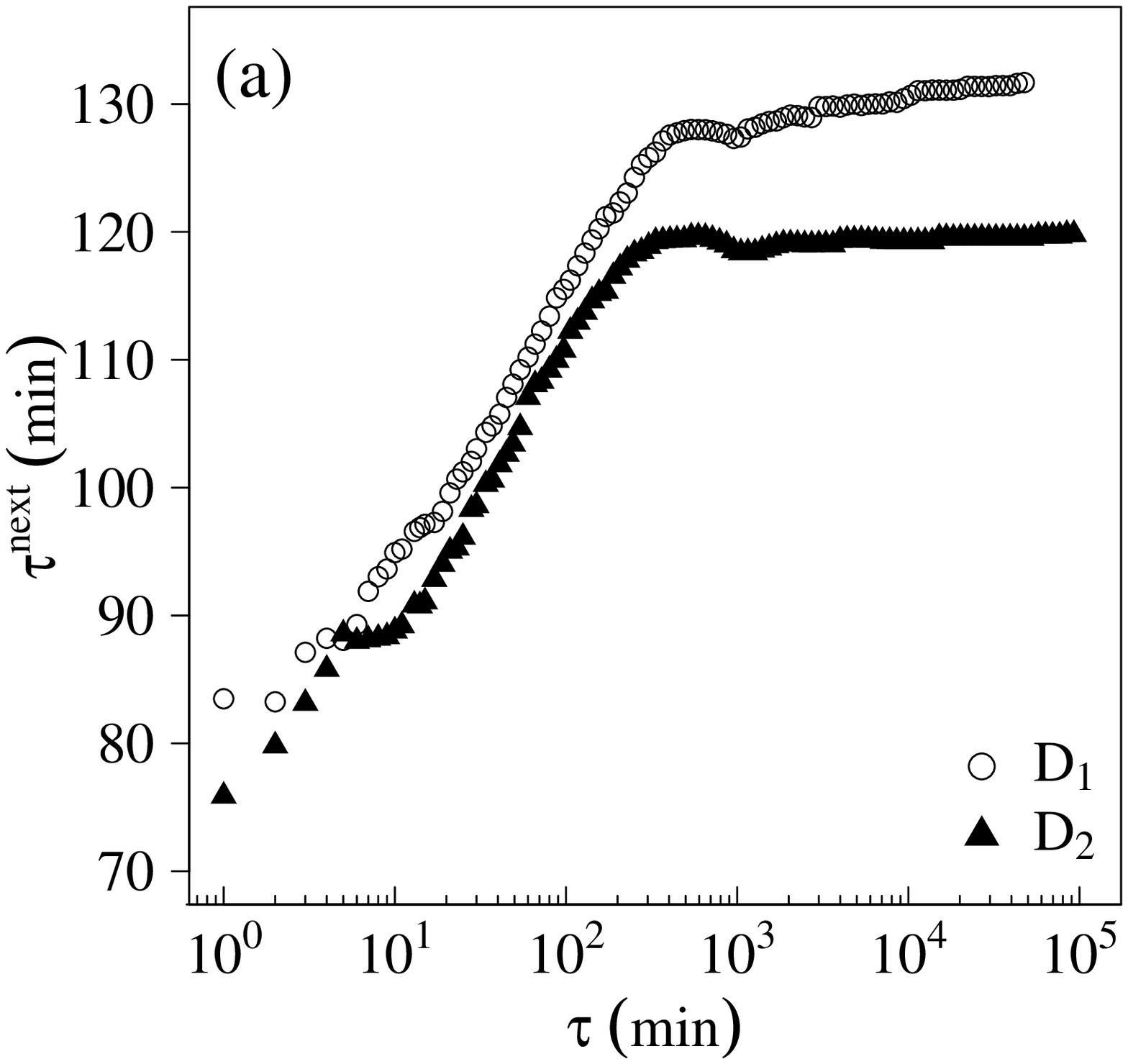}
\includegraphics[width=6cm]{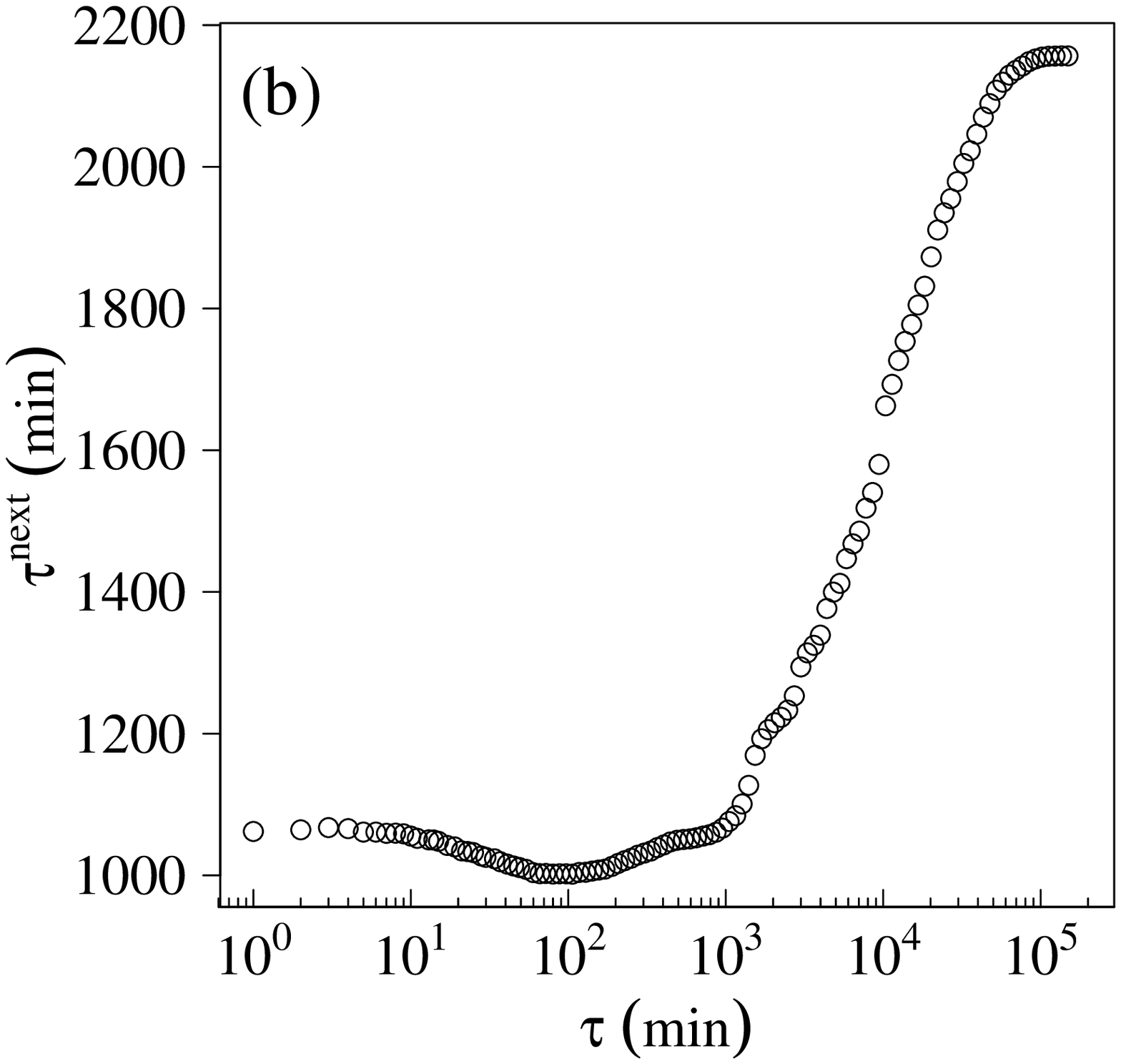}
\hspace*{-1cm}
\includegraphics[width=6cm]{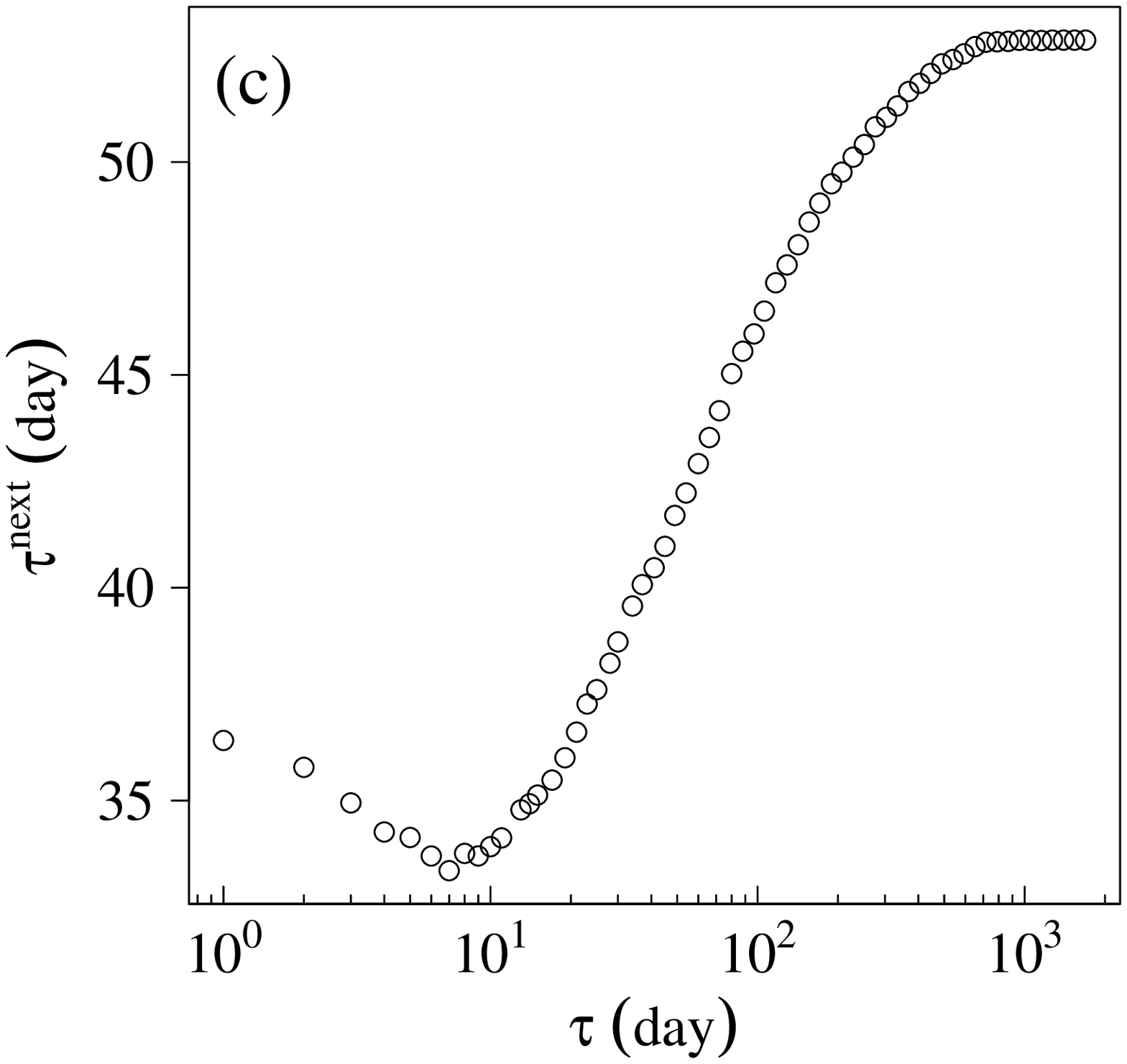}
\includegraphics[width=6cm]{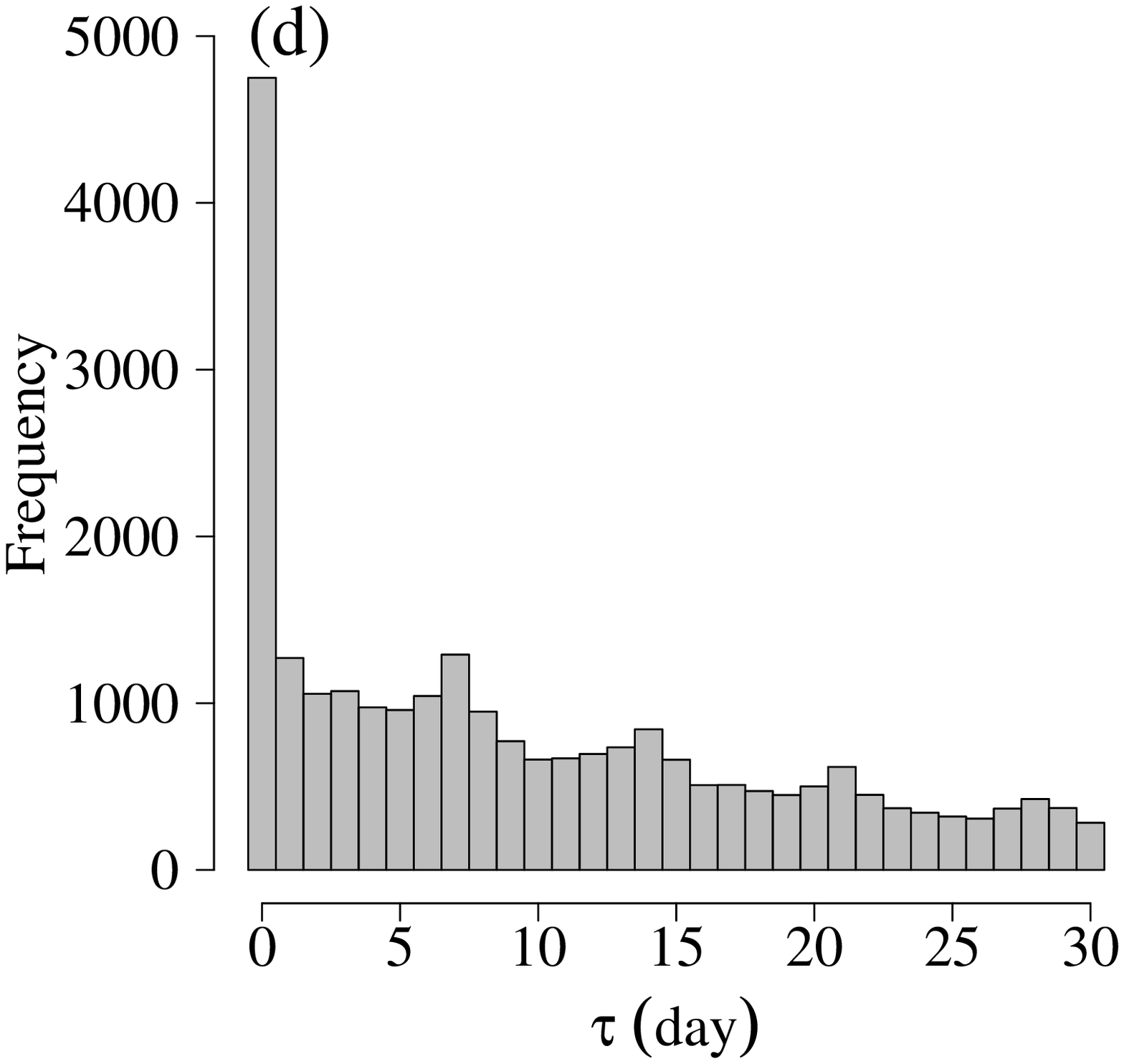}
\caption{Conditional mean IET defined by \EQ\eqref{eq:tau next} for
(a) conversation event sequences \cite{Takaguchi2011PRX},
(b) email logs \cite{Ebel2002PRE_email}, and (c)
purchase of sexual escorts \cite{Rocha2010PNAS}. (d) Histogram of the IET
for the data shown in (c).}
\label{fig:conditional IET}
\end{center}
\end{figure}

The values of $\tau^{\rm next}$ are plotted against $\tau$
in \FIGS\ref{fig:conditional IET}(a), \ref{fig:conditional IET}(b),
and \ref{fig:conditional IET}(c) for the
conversation sequences used in
\FIG\ref{fig:IET dist}, times of
email sending and receiving in a university \cite{Ebel2002PRE_email},
and times of online sexual escorts by male
individuals \cite{Rocha2010PNAS}, respectively.
We remark that long-tailed IET distributions are known
for the email \cite{Barabasi2005Nature,VazquezA2006PRE_burst}
%
%
and sexual escort \cite{Rocha2010PNAS} data sets.
%
%
The conditional mean IET $\tau^{\rm next}$ increases with
$\tau$ in \FIGS\ref{fig:conditional
  IET}(a) and \ref{fig:conditional IET}(b). 
Therefore, adjacent IETs are positively correlated.
In \FIG\ref{fig:conditional IET}(c), $\tau^{\rm next}$
decreases with $\tau$ for $\tau\le 7$ and increases with $\tau$ for $\tau\ge
7$. Figure~\ref{fig:conditional IET}(c)
suggests that those who have bought an escort tend to avoid
buying a next escort within a week.
This is directly shown in \FIG\ref{fig:conditional IET}(d), which shows the IET distribution. However, adjacent IETs for the sexual escort data
are positively correlated on a longer time scale
(\FIG~\ref{fig:conditional IET}(c)).

In the discrete time model proposed in
\cite{HanZhouWang2008NJP}, the probability of an event occurrence
decreases if
events occurred too frequently in the recent past and increases if the
time since the last event becomes long. Such a mechanism may generate positive IET correlation.

\subsection{Self-excitatory Stochastic Processes}

An alternative mechanism that yields positive IET correlation
is self-excitation. The idea is that once an individual talks with
somebody,
the individual is excited to talk with somebody with a higher rate.
Malmgren and coworkers developed such models and applied to data 
\cite{Malmgren2008PNAS,Malmgren2009Science}.  

In the cascading
nonhomogeneous Poisson process proposed in \cite{Malmgren2008PNAS},
the authors assumed that the primary process is an inhomogeneous
Poisson process with a periodic event rate.
An event generated from the primary process is
assumed to elevate the system to the active state and
trigger cascades of activity. In other words,
after a trigger event, a burst of events may ensue as a result of
the Poisson process with a rate that is larger than the base rate of
the primary process. The entire recording period is divided into alternately appearing
intervals of the active state with a high event rate and the normal state with a low event rate by 
an adjustment of the position and number of intervals to yield a good fit to the data.
As a result, the number of events contained in a burst is shown to
obey an approximate exponential distribution (also see \cite{Karsai2012SciRep}, which shows that the number of events in a burst obeys a power law distribution; the definition of burst is different in the two papers).
With a circadian and
weekly rate modulation, the cascading nonhomogeneous Poisson process
is capable of producing the long-tailed IET distributions observed in the data.

Their model has many parameters to be estimated.
This is common to their another model proposed in
\cite{Malmgren2009Science}. In \cite{Malmgren2009Science},
letter writing activity of each renowned individual is fitted by a cascading Poisson process model. The time unit is set to a day. The two parameters, i.e., the
base event rate and tendency to write an additional letter within a time unit, are estimated on the basis of the data. Because the different parameter values are assumed for different sections of the data,
the number of the parameters in the model can be large.
In the case of the letter correspondence by Einstein, data are collected over 54 years, and the two parameters are estimated for each year. Therefore,
there are 108 parameters.

These models \cite{Malmgren2008PNAS,Malmgren2009Science} are quite successful in capturing
properties of the real event sequences. Nevertheless, it may be also fruitful to consider a much simpler self-excitatory model as a complementary approach to capture the origins of bursts (\SEC\ref{sub:long tail IET}) and IET correlation (\SEC\ref{sub:IET correlation}) inherent in human behavior.

A simple two state model in which normal and excited
states are assumed is proposed in \cite{Karsai2012SciRep}. The model is not a
hidden Markov model because the probability of staying in the excited
state becomes large as the number of events that have already occurred
in the current burst increases.  The model reproduces properties of the original data such as the power-law
IET distribution and autocorrelation function.
However, statistical methods to estimate the model parameters from the
data were not presented \cite{Karsai2012SciRep}.

\subsection{Our Approach: Hawkes Process}

In this chapter, we fit the self-excitatory
point process model called the Hawkes process
\cite{Hawkes1971Biom,Hawkes1971JRSSB,Hawkes1974JAP,Verejones1970JRSSB} to the data recorded in
company offices \cite{Wakisaka2009IEEE,Yano2009Hitachi,Takaguchi2011PRX,Takaguchi2012NJP}
(also see \FIG\ref{fig:IET dist} and \FIG\ref{fig:conditional IET}(a)).
A main benefit for using the Hawkes process is that it contains a
small number of parameters and is mathematically tractable;
the maximum likelihood (ML) method for inferring parameter values
is established for some important special cases \cite{Ozaki1979AISM}.

This chapter is organized as follows.
In \SEC\ref{sec:def Hawkes}, we introduce the Hawkes model and 
recapitulate its basic mathematical properties. In \SEC\ref{sec:IET Hawkes}, we numerically investigate properties of event sequences generated by
the Hawkes process. In \SEC\ref{sec:results},
we carry out the ML estimation of the parameters of the Hawkes model and
compare the data and the estimated model. In \SEC\ref{sec:discussion}, we discuss the results, with an emphasis on the limitation and possible extensions of the Hawkes model for better describing human data. Mathematical details are delegated to two Appendices.

\section{Hawkes Process}\label{sec:def Hawkes}

The Hawkes process is a self-exciting point process model that is analytically tractable.
 It is an inhomogeneous
Poisson process in which the instantaneous event rate depends on the
history of the time series of events. It is not a renewal process.
The event rate at time $t$, denoted by
$\lambda(t)$ is given by
\begin{equation}
\lambda(t) = \nu + \sum_{i,t_i\le t}\phi(t-t_i),
\end{equation}
where $t_i$ is the time of the $i$th event, and $\phi(t)$ is the
memory kernel, i.e., the additional rate incurred by an event. The
causality implies $\phi(t)=0$ ($t<0$).

The Hawkes process has been used for modeling, for example,
seismological data \cite{Verejones1970JRSSB,Ogata1999PAG}, 
 video viewing activities \cite{Crane2008PNAS,Mitchell2010JPA}, neural spike trains \cite{Pernice2011PlosComputBiol}, and
genomic data \cite{ReynaudBouret2010AnnStat}.
For example, in \cite{Crane2008PNAS}, time series of views of different videos on YouTube
were categorized into three classes, which were characterized by different $\phi(t)$ and different 
time-dependent versions of $\nu$.
The Hawkes process has also been used to construct 
a method to estimate the structure of neural networks from given spike trains \cite{Dahlhaus1997JNSMethods}, analyze auto and cross correlation in data recorded from mouse retina
\cite{Krumin2010FCN}, and understand the
correlation between the activities of different neurons in
pulse-coupled model networks of excitatory and inhibitory neurons
\cite{Pernice2011PlosComputBiol}.
In \cite{ReynaudBouret2010AnnStat},
the Hawkes process is used to model stochastic occurrences of
specific genes 
 on DNA sequences. The method to estimate a piecewise linear $\phi(t)$ based on the least square
error was presented.

\begin{figure}
\begin{center}
\includegraphics[width=12cm]{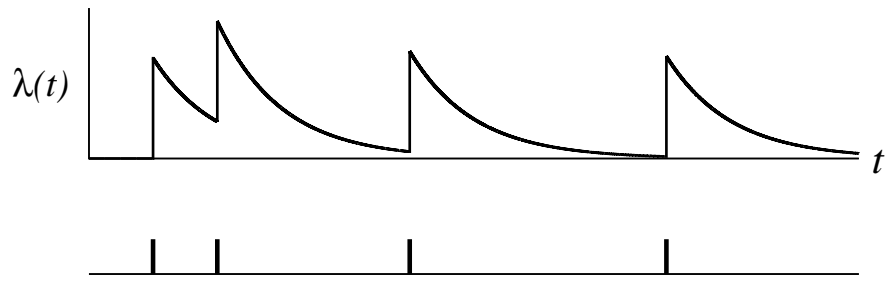}
\caption{Example time course of event rate $\lambda(t)$ and the corresponding event sequence.}
\label{fig:example rate Hawkes}
\end{center}
\end{figure}

Depending on applications, the memory kernel
$\phi(t)$ has been assumed to be
a hyperbolic (i.e., power law) function \cite{Crane2008PNAS}
or a superposition of the gamma function \cite{Ogata1999PAG}.
Nevertheless, in the present work, we simply set
\begin{equation}
\phi(t)=\alpha e^{-\beta t} (t\ge 0)
\label{eq:phi def exponential}
\end{equation}
for the following reasons. First, it allows the
ML estimation of the parameters $\alpha$, $\beta$, and $\nu$
\cite{Ozaki1979AISM}. Second, the Hawkes process with
\EQ\eqref{eq:phi def exponential} has a small number of
parameters as compared to competitive models with self excitation
\cite{Malmgren2008PNAS,Malmgren2009Science,OgataAkaike1982JRSSB,Ogata1999PAG}.
It should be noted that \EQ\eqref{eq:phi def exponential} indicates that the self-exciting effect of an event decays in time. It is contrasted with a previous model in which the self-exciting effect is constant for some time and then the event rate returns to the basal rate \cite{Malmgren2008PNAS}. An example time course of the event rate $\lambda(t)$ and the corresponding event sequence are shown in \FIG\ref{fig:example rate Hawkes}.

We define cluster of events as the set of events that are triggered by a single
event occurring at the basal rate $\nu$. In other words, all the events in a cluster are descendants
of the trigger event. The expected cluster size
is given by \cite{Hawkes1971Biom,Verejones1970JRSSB}
\begin{equation}
c=\int^{\infty}_0 \phi(t)dt = \frac{1}{1-\frac{\alpha}{\beta}},
\label{eq:csize}
\end{equation}
and the stationary event rate is given by
\begin{equation}
\overline{\lambda}=c\nu = \frac{\nu}{1-\frac{\alpha}{\beta}}.
\label{eq:overline(lambda)}
\end{equation}
The convergence of the event rate requires $\alpha<\beta$.

\section{Numerical Results for Statistics of IET}\label{sec:IET Hawkes}

In this section, we numerically examine basic properties of the Hawkes process with the
exponential memory kernel.
To quantify the broadness of the IET distribution, we measure the
coefficient of variation (CV), defined as the standard deviation of
the IET divided by the mean of the IET as follows:
\begin{equation}
\mbox{CV } = \frac{\sqrt{\sum_{i=1}^N (\tau_i-\left<\tau\right>)^2/N}}
{\left<\tau\right>},
\end{equation}
where $N$ is the number of IETs in a given sequence
and $\left<\tau\right> \equiv \sum_{i=1}^N \tau_i/N$ is the mean IET.
It should be noted that
$\left<\tau\right>$ in the limit $N\to\infty$ is equal to
$1/\overline{\lambda}=(1-\alpha/\beta)/\nu$. The Poisson process
yields CV $=1$. 

We also measure the correlation coefficient for the IET \cite{Goh2008EPL} defined as
\begin{equation}
\frac{\sum_{i=1}^{N-1} (\tau_i-\left<\tau\right>)(\tau_{i+1}-\left<\tau\right>)/(N-1)}
{\sum_{i=1}^N (\tau_i-\left<\tau\right>)^2/N}.
\end{equation}

The Hawkes process is invariant under
the following rescaling of the time and parameter values:
$Ct = t^{\prime}$,
$\alpha = C\alpha^{\prime}$,
$\beta = C\beta^{\prime}$, and
$\nu = C\nu^{\prime}$, where $C>0$ is a constant.
Therefore, we normalize the time by setting
$\nu=1$ and vary $\alpha$ and $\beta$. The values of CV, IET correlation, and
mean cluster size $c$ are invariant under this rescaling.
For a given pair of $\alpha$ and
$\beta$ values, we generate 
a time series with $2\times 10^5$ events using the method described in \cite{Ogata1981IEEEIT} and calculate the statistics of the IET.

\begin{figure}
\begin{center}
\includegraphics[width=6cm]{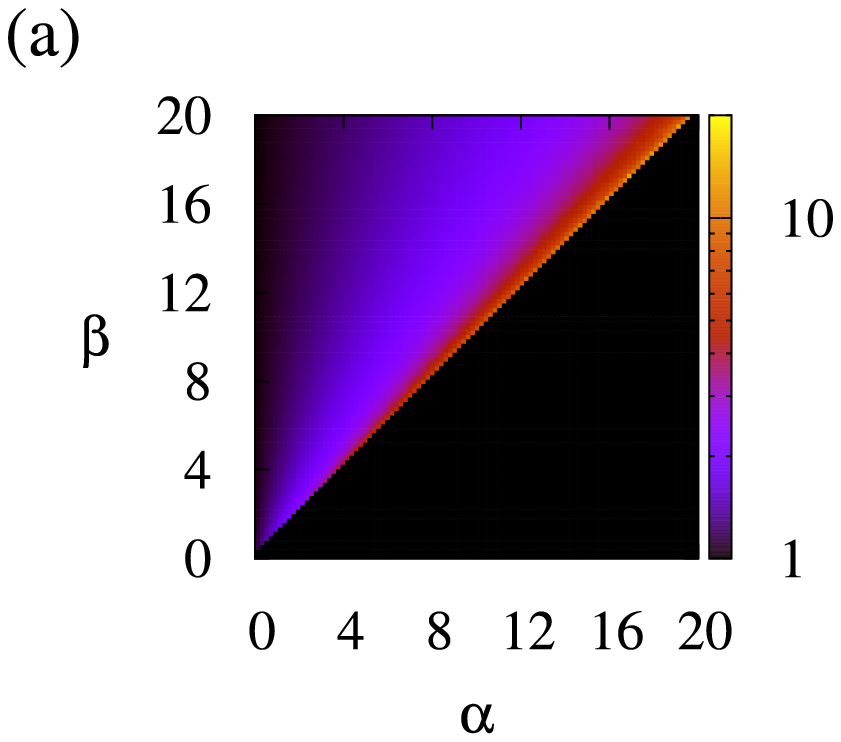}
\includegraphics[width=6cm]{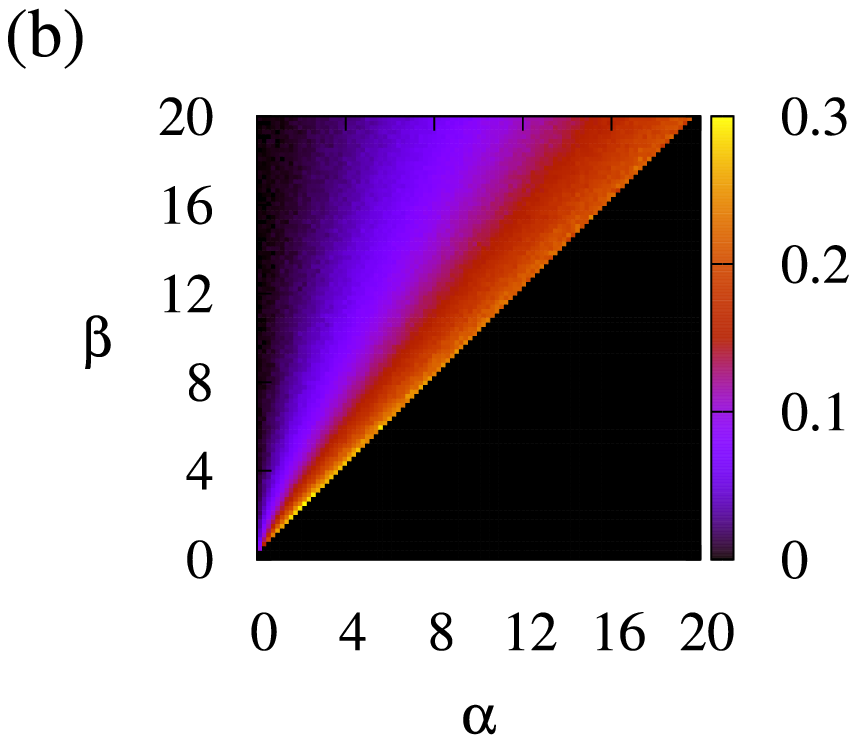}
\includegraphics[width=6cm]{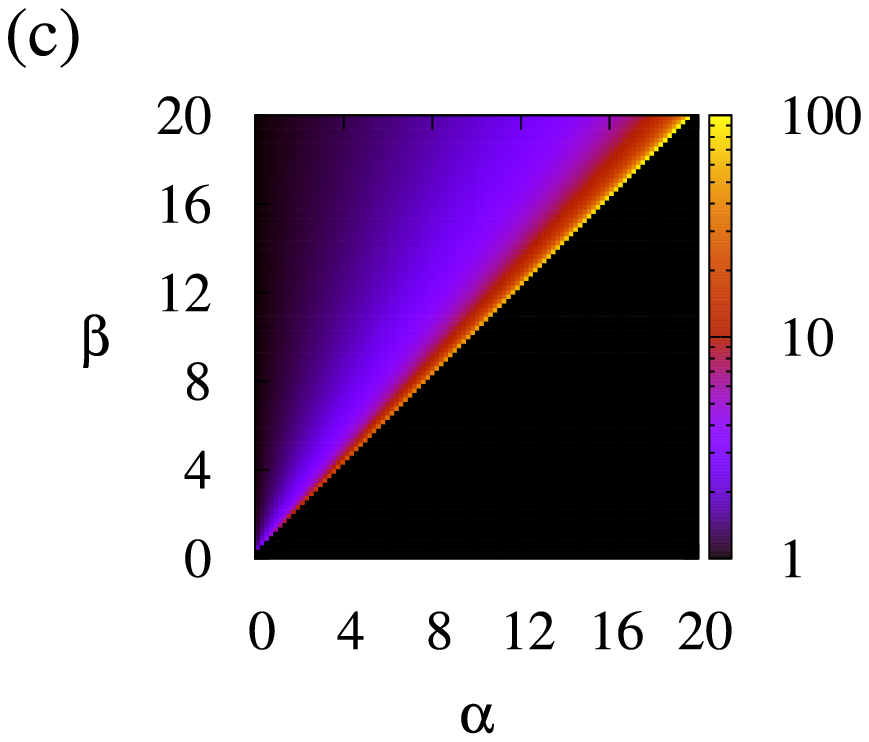}
\caption{Statistics of the IET obtained from the Hawkes process. (a) CV, (b) IET correlation, and (c) mean cluster size $c$ for various values of $\alpha$ and $\beta$.}
\label{fig:stat IET Hawkes p=0}
\end{center}
\end{figure}

The values of CV, IET correlation, and $c$ (\EQ\eqref{eq:csize}) for various
$\alpha$ and $\beta$ values are shown in
\FIG\ref{fig:stat IET Hawkes p=0}(a), \FIG\ref{fig:stat IET Hawkes p=0}(b), and
\FIG\ref{fig:stat IET Hawkes p=0}(c), respectively.
Although we can more theoretically calculate CV
using the expression of the IET distribution \cite{Hawkes1974JAP}
(also see Appendix 1 for details),
it is numerically demanding to do so. Therefore,
we resorted to direct numerical simulations.
The data are present only in the
region $\alpha<\beta$, where the Hawkes process does not explode.

\begin{figure}
\begin{center}
\includegraphics[width=6cm]{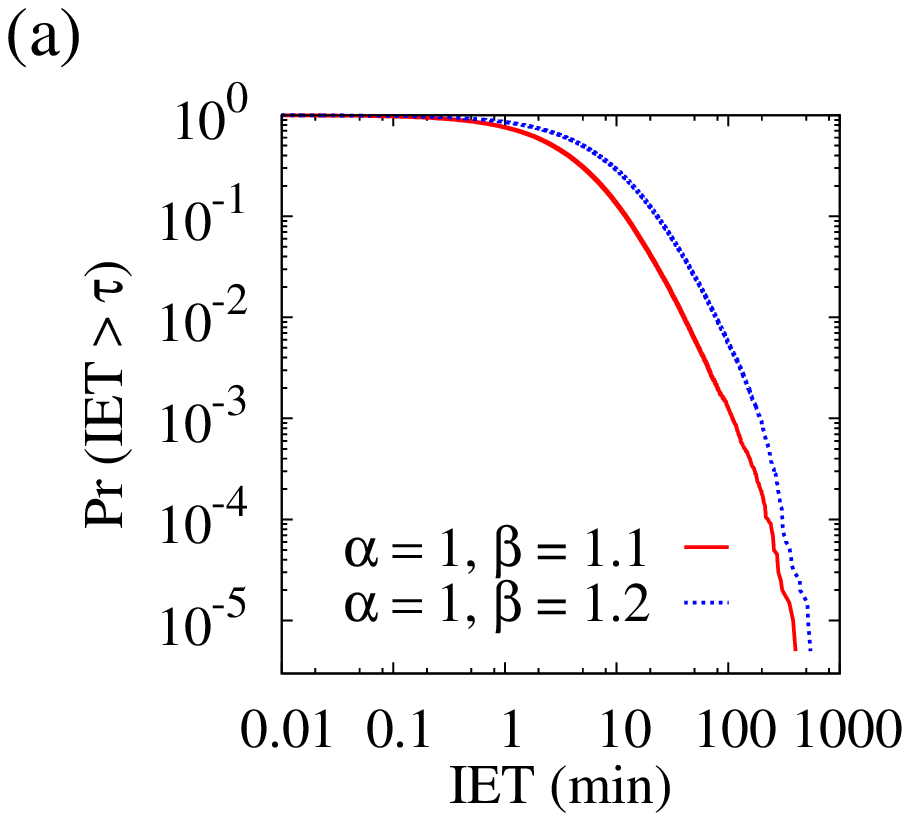}
\includegraphics[width=6cm]{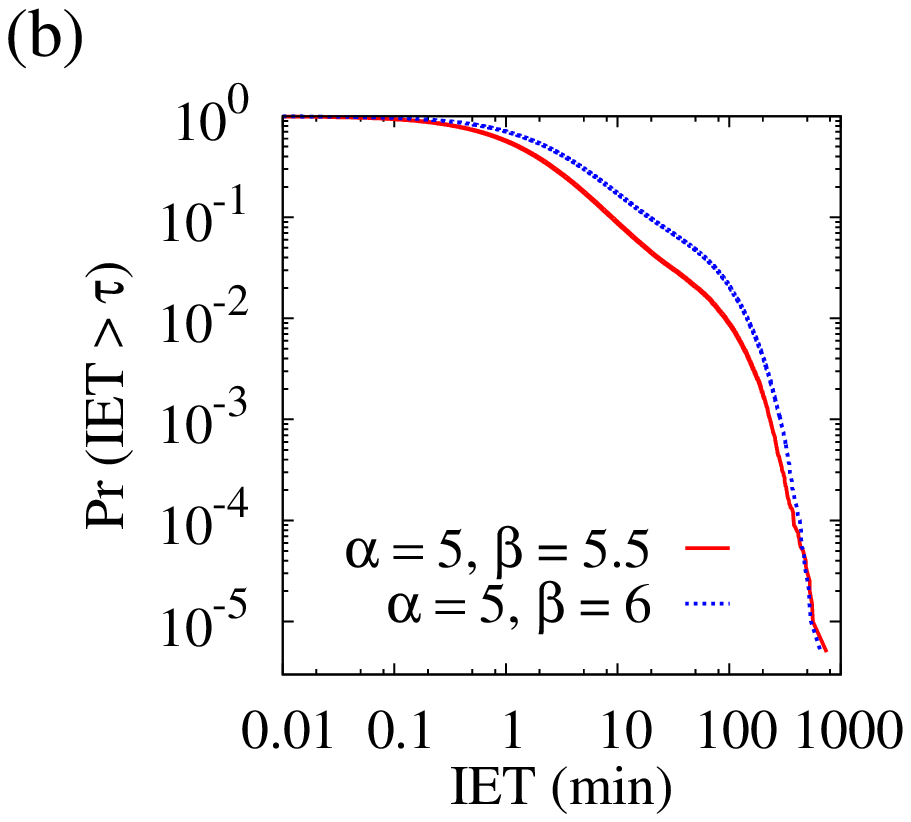}
\includegraphics[width=6cm]{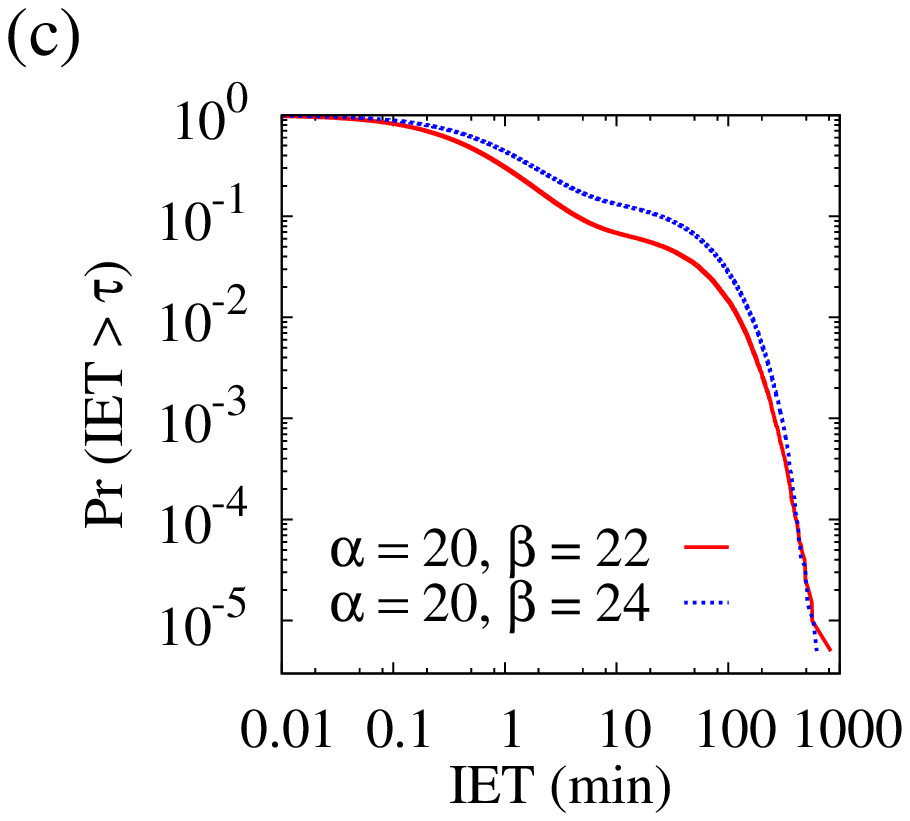}
\caption{Survivor functions of the IET for the Hawkes process with different values of $\alpha$ and $\beta$.}
\label{fig:IET dist examples}
\end{center}
\end{figure}

Figure~\ref{fig:stat IET Hawkes p=0}(a) indicates that the Hawkes process generates
a wide range of CV. A large value of $\alpha/\beta (<1)$ yields a
large CV value. 
This is the case for both small and large $\alpha$ values.
In \FIG\ref{fig:IET dist examples}, the survival function
of the
IET on the basis of 
$2\times 10^5$ events is compared for different $\alpha$ and $\beta$ values that satisfy $\beta=1.1\alpha$ or $1.2\alpha$.
Although
the CV values are large, the IET distributions are consistently different from
power law distributions. In particular, the IET distribution
seems to be a superposition of multiple distributions with different time scales
when $\alpha$ is large (\FIG\ref{fig:IET dist examples}(c)).
It should be noted that we assumed the exponential, not long-tailed,
memory kernel (\EQ\eqref{eq:phi def exponential}).

Figure~\ref{fig:stat IET Hawkes p=0}(b) indicates that a
large $\alpha/\beta$ value also yields a large IET correlation.  Once
the event rate increases because of recent occurrences of other
events, the following IET tends to be small. Therefore, strong
self-excitation in the model (i.e., large $\alpha/\beta$) is
considered to cause large IET correlation. The strength of self-excitation can be also quantified by $c$.
Figure~\ref{fig:stat IET Hawkes p=0}(c) indicates that
a large $\alpha/\beta$ tends to yield a large $c$.

Figures~\ref{fig:stat IET Hawkes p=0}(a), \ref{fig:stat IET Hawkes p=0}(b), and \ref{fig:stat IET Hawkes p=0}(c)
look similar,
suggesting that the three quantities are
correlated with each other.

\section{Fitting the Hawkes Process to the Data}\label{sec:results}

\subsection{Data Sets}\label{sub:data sets}

We analyze two data sets $D_1$ and $D_2$ of face-to-face interaction
logs obtained from different company offices in Japan.  World Signal
Center, Hitachi, Ltd., Japan collected the data using the Business
Microscope system developed by Hitachi, Ltd., Japan.  For technical
details concerning the data collection, see
\cite{Wakisaka2009IEEE,Yano2009Hitachi,Takaguchi2011PRX}.
We previously analyzed the data using different methods \cite{Takaguchi2011PRX,Takaguchi2012NJP}.
 Data sets
$D_1$ and $D_2$ consist of recordings from 163 individuals for 73
days and 211 individuals for 120 days, respectively. 
The two
individuals are defined to be involved in a conversation event, simply
called the event, if their modules exchange the IDs at least once in a
minute. The module has other types of data that we do not use in the present study, such as the list of
conversation partners and the duration of each event.
In total, $D_1$ and $D_2$ contain 51,879
and 125,345 events, respectively.

\subsection{Results of Fitting}

For the entire sequence of event times obtained for each individual,
we carry out the ML estimation of the parameters of the Hawkes process with the exponential memory
kernel. It should be noted that we use the information about event times and not the duration of events or the partners' IDs.
We slightly modify the ML method developed in \cite{Ozaki1979AISM}
(see Appendix 2 for details).

The modification is concerned with the treatment of the data during the night.
Our data are nonstationary owing to the circadian and weekly
rhythms. Therefore, direct application of the Hawkes process, which is a stationary point process, is invalid. In the previous literature in which different models are investigated, these rhythms are explicitly modeled \cite{Gonzalez2008Nature,Malmgren2008PNAS} or treated by dynamically changing the time scale according to the event rate
\cite{Jo2012NewJPhys}. In contrast,
 we omit the night part of the data from the analysis because our data are collected in company offices and therefore there is no event from late in the night through early in the morning.

In both data sets $D_1$ and $D_2$, there is nobody in the office
between four and six in the morning. Accordingly, we can partition the
data into workdays without ambiguity. For each individual, we discard the
workdays that contain less than 40 conversation events. We call a
workday containing at least 40 events the valid day. Then, we define the
first event in each valid day as trigger event and set $t=0$. The
following events on the same valid day are interpreted to be generated
from the Hawkes process.  The time of the last event denoted by $t_{\rm last}$
(denoted by $t_{N_d}^d$ in Appendix 2) is
defined to be the end time of the valid day; it is necessary to specify
$t_{\rm last}$ to apply the ML method (Appendix 2).
The value of $t_{\rm last}$ depends on individuals even on the same day.
The individual may stay in the office
for a considerable amount of time after $t=t_{\rm last}$ before 
leaving the office.
This implies that the individual does not have conversations
with others remaining in the office between $t=t_{\rm last}$ and the time when
the individual leaves the office.
If this is the case, the fact that
this individual does not have events for $t>t_{\rm last}$ may affect the
ML estimators.  Nevertheless, we neglect this point. Finally, we obtain
the likelihood of the series of events for an individual
by multiplying the likelihood for all the valid days.

We apply the ML method to the individuals that possess at least
300 valid IETs (i.e., IETs derived from the valid days)
during the entire period.  This thresholding leaves 63
individuals in $D_1$ and 148 individuals in $D_2$.
We also exclude one individual 
in $D_1$ because the ML method does not converge for this individual.

\begin{figure}
\begin{center}
\includegraphics[width=6cm]{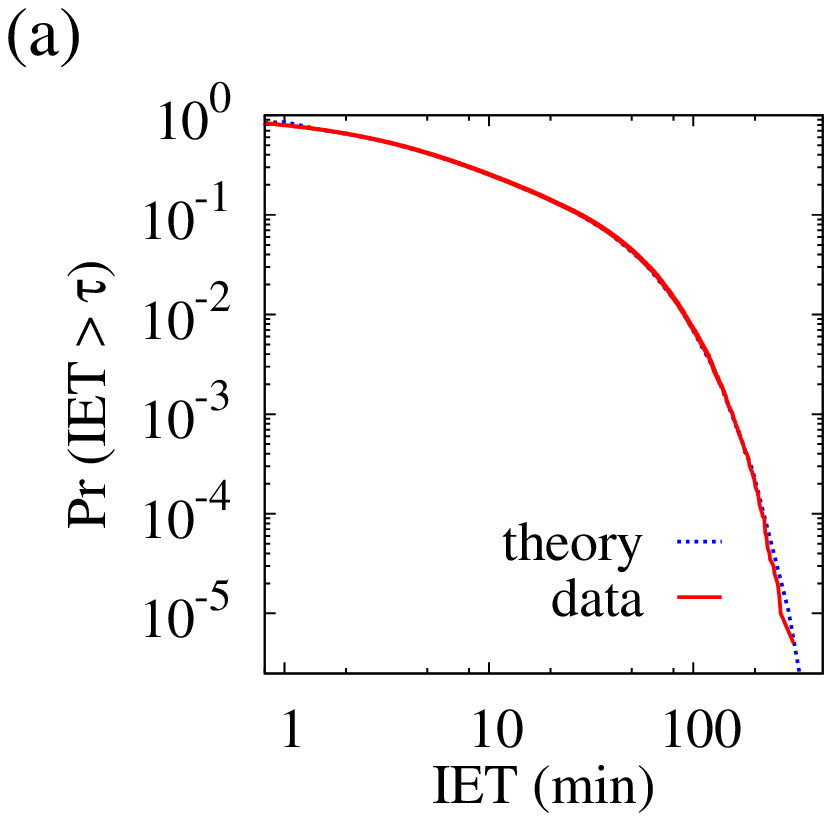}
\includegraphics[width=6cm]{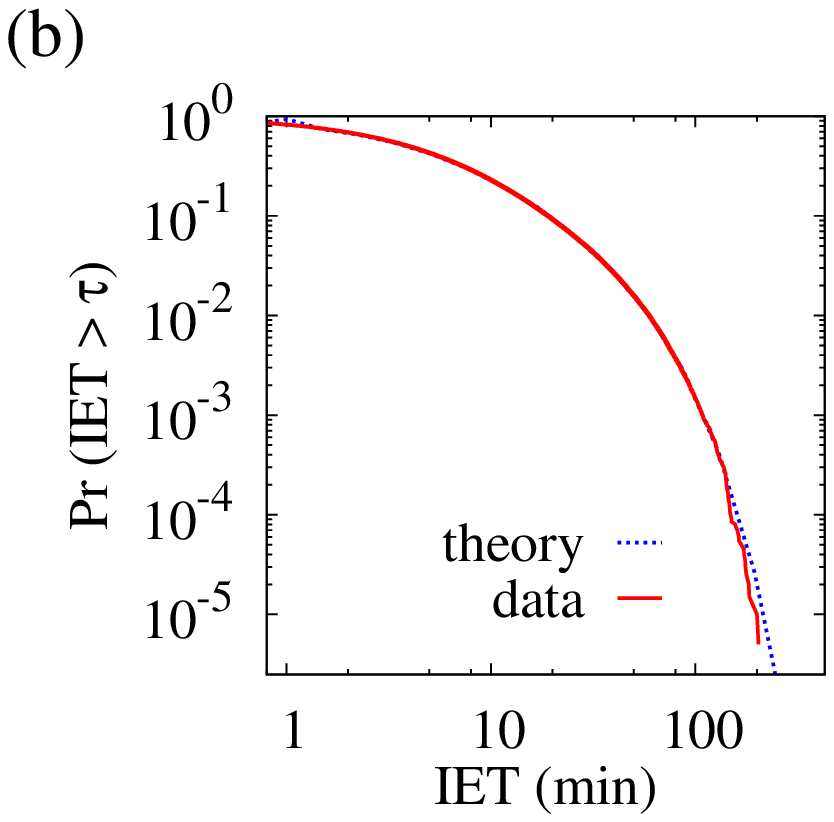}
\caption{Survivor functions of the IET for two individuals. (a) Results for an individual in $D_1$,
who has 1694 valid IETs during the recording period. The ML estimators are given by $\alpha=4.91$, $\beta=7.89$, and $\nu=2.18$.
(b) Results for an individual in $D_1$, who has 1765 valid IETs. The ML estimators are given by $\alpha=2.45$, $\beta=3.86$, and $\nu=2.77$.}
\label{fig:IET fit}
\end{center}
\end{figure}

The survivor function of the IET is compared between the data and
the estimated Hawkes process in \FIG\ref{fig:IET fit}. The comparison
is made for an individual in $D_1$ (\FIG\ref{fig:IET fit}(a)) and
an individual in $D_2$ (\FIG\ref{fig:IET fit}(b)). 
We calculated the IET
distribution for the estimated model using the theoretical method
\cite{Hawkes1974JAP} (Appendix 1).
The agreement between the IET distributions of the data and 
the estimated model is excellent.

\begin{figure}
\begin{center}
\includegraphics[width=6cm]{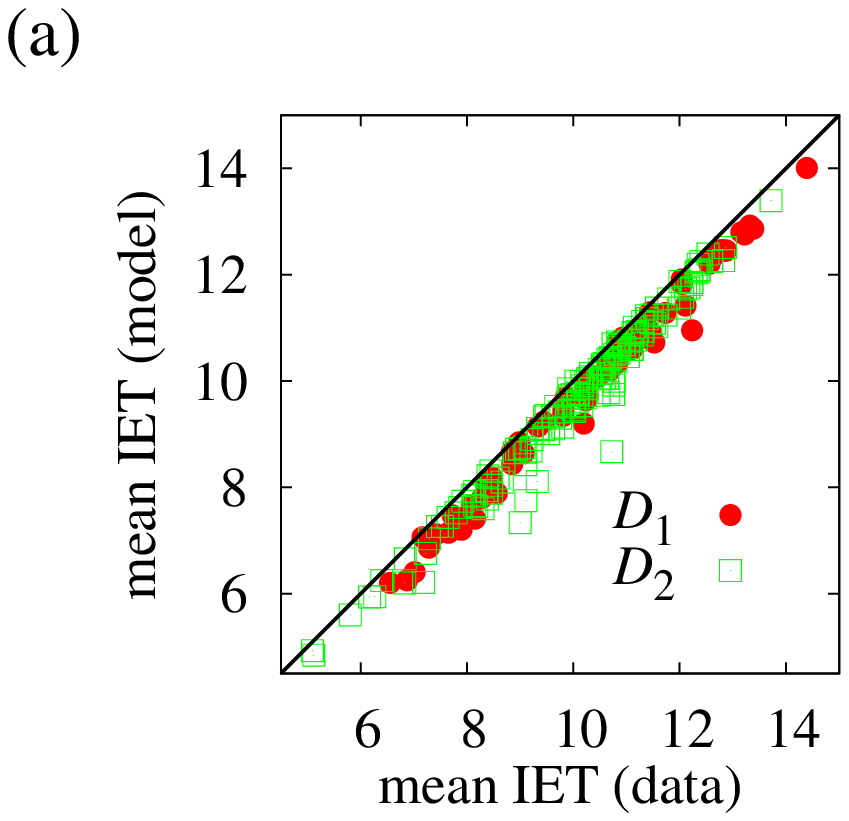}
\includegraphics[width=6cm]{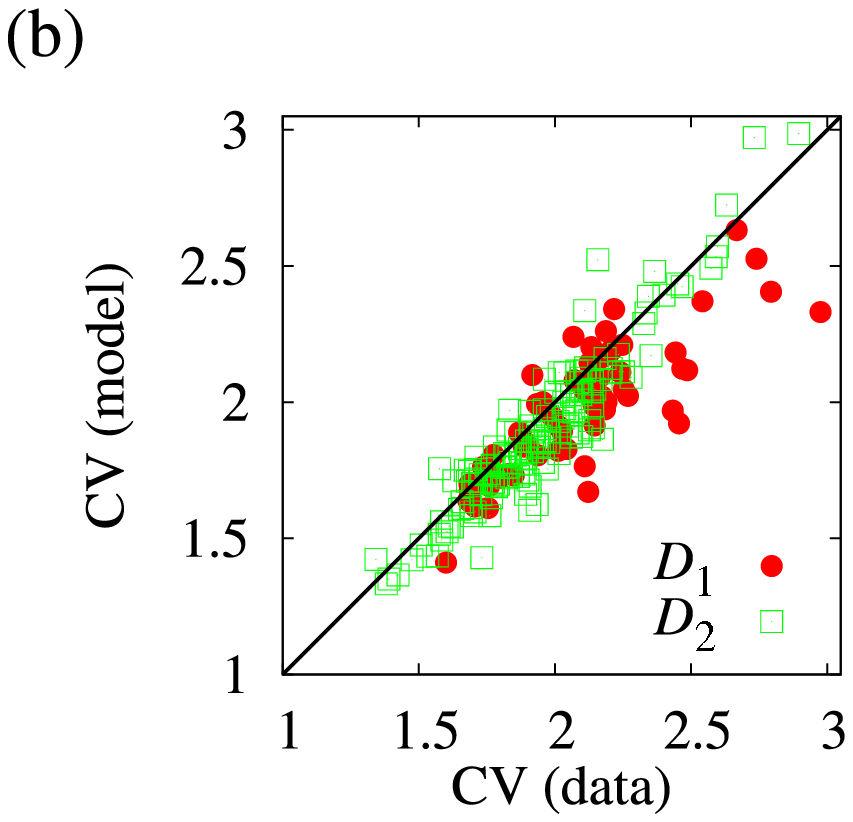}
\includegraphics[width=6cm]{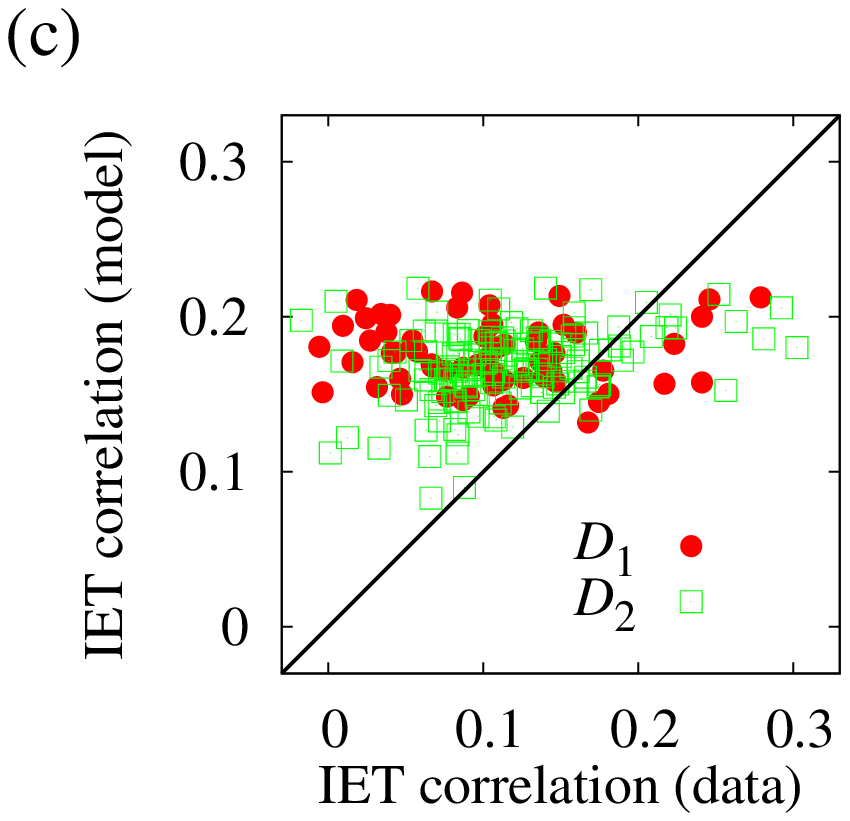}
\caption{Comparison between the data and the estimated model. (a) mean IET, (b) CV, and (c) IET correlation. Each data point corresponds to one valid individual.
The mean IET, CV, and IET correlation for the data are calculated on the basis of
the days containing at least 40 events.}
\label{fig:compare Hawkes}
\end{center}
\end{figure}

To assess the quality of the fit at a population level,
we compare three statistics of the IET between the data and model for different individuals.
The relationship between the mean IET obtained from the data and that
obtained from the estimated model, i.e.,
$1/\overline{\lambda}=(1-\alpha/\beta)/\nu$, is shown in
\FIG\ref{fig:compare Hawkes}(a).
For different individuals in both
data sets, the mean IET is close between the data and the
model. The Pearson correlation coefficient between the data and model
are equal to 0.993 and 0.986 for $D_1$ and
$D_2$, respectively.
However, the Hawkes process slightly underestimates the mean IET.  

The CV
values for the data and the estimated model are compared in
\FIG\ref{fig:compare Hawkes}(b). We calculated the CV values for the
estimated model on the basis of $2\times
10^5$ events that we obtained by simulating the Hawkes process with the
ML estimators $\alpha$, $\beta$, and $\nu$.
Although the CV can be theoretically calculated using the ML estimators
(Appendix 1), we avoided doing so because the theoretical method is
computationally too costly to be applied to all the individuals.
Roughly speaking, the CV values obtained from the model are close to
those of the data.
The Pearson correlation coefficient between the data and model
are equal to 0.832 and 0.936 for $D_1$ and
$D_2$, respectively.

The IET correlation of the data and that for the estimated model are
compared in \FIG\ref{fig:compare Hawkes}(c).
We calculated the IET correlation for
the estimated model by direct numerical simulations, as in the case of
the CV. Figure~\ref{fig:compare Hawkes}(c) indicates that 
the Hawkes process does not reproduce
the IET correlation for most individuals. The IET correlation for the
estimated model is distributed in a much narrower range
than that of the data. This is consistent with the finding that
the CV and the IET correlation are positively correlated in the Hawkes
process (\SEC\ref{sec:IET Hawkes}). Because most individuals have the CV
values larger than unity (\FIG\ref{fig:compare Hawkes}(b)), the
estimate of the IET correlation obtained by the model tends to be
positive regardless of the estimated values of $\alpha$, $\beta$, and $\nu$.
Figure~\ref{fig:compare Hawkes}(c) suggests that
the Hawkes process with the exponential memory kernel is incapable of
approximating the real data in terms of the IET correlation.

\section{Discussion}\label{sec:discussion}

We analyzed properties of the IET generated by the Hawkes process with an exponential
memory kernel and then fitted the model to the face-to-face interaction logs obtained from company offices. The model successfully reproduced the data in terms of the IET distribution.
However, the model does not
explain the behavior of the IET correlation in the data.

\begin{figure}
\begin{center}
\includegraphics[width=6cm]{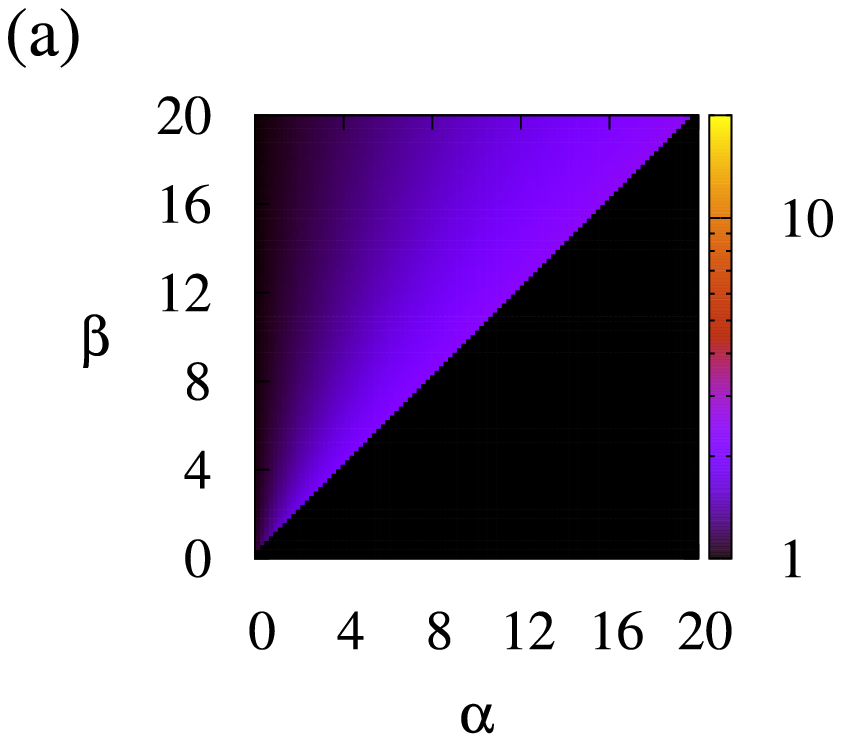}
\includegraphics[width=6cm]{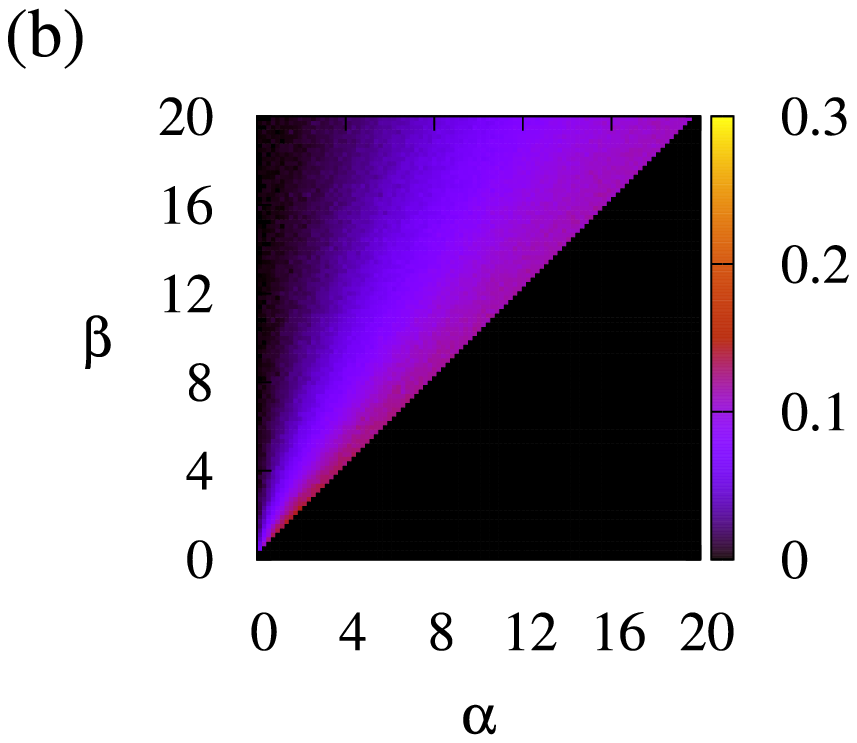}
\includegraphics[width=6cm]{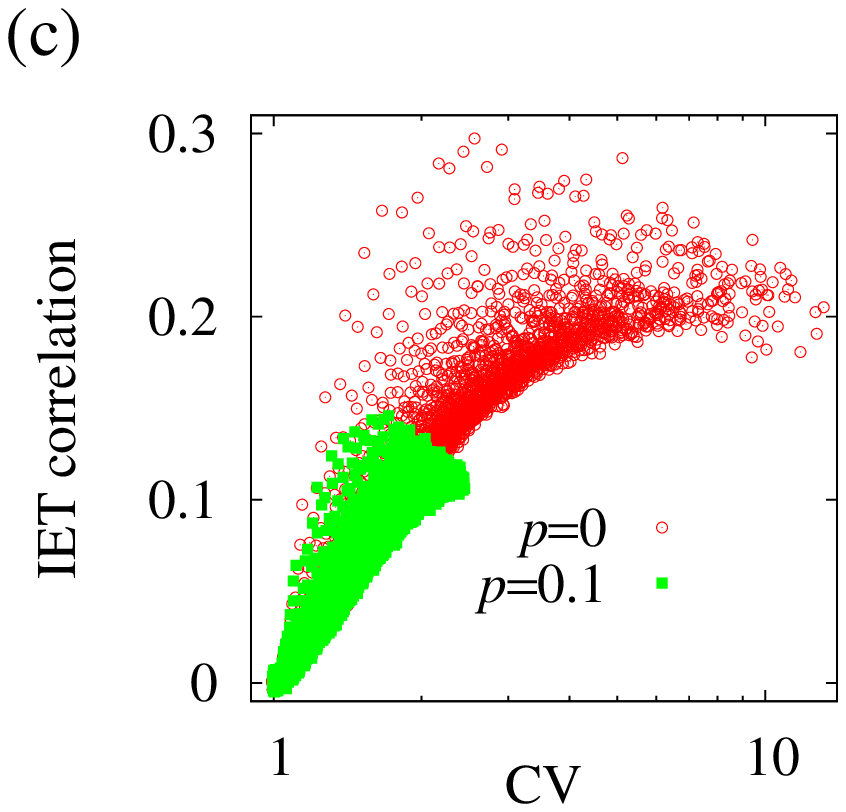}
\caption{Results for the modified Hawkes process with $\nu=1$. (a) CV with $p=0.1$. (b) IET correlation with $p=0.1$. (c) Relationship between the CV and IET correlation with $p=0$ and $p=0.1$.}
\label{fig:stat IET Hawkes p=0.1}
\end{center}
\end{figure}

This limitation
may be because the effect of self-excitation is too strong in the
Hawkes process; the event rate can be very large after a burst of
events. To examine this issue, we carry out additional numerical simulations
using a modified Hawkes model. We modify the model such that after
each event that would increase the event rate by $\phi(0)$ in the
original Hawkes process, we reset the event rate to the basal value
$\nu$ with probability $p$. The original Hawkes process corresponds to
$p=0$. The CV and IET correlation for $p=0.1$ and various values of
$\alpha$ and $\beta$ are shown in \FIGS\ref{fig:stat IET Hawkes
  p=0.1}(a) and \ref{fig:stat IET Hawkes p=0.1}(b), respectively.  The
values of the CV and IET correlation for $p=0.1$ are much smaller than
those for $p=0$ (\FIGS\ref{fig:stat IET Hawkes p=0}(a) and
\ref{fig:stat IET Hawkes p=0}(b)).  This is because a burst, which
increases the CV and IET correlation in the Hawkes model,
is forced to terminate
with probability $p$ after each event in the modified model.  The CV
and IET correlation values for $(\alpha, \beta) = (0.2i, 0.2j)$, where
$0\le i<j<100$ are plotted in \FIG\ref{fig:stat IET Hawkes
  p=0}(c). For comparison, the corresponding results for $p=0$ on the basis of the
data used in \FIGS\ref{fig:stat IET Hawkes p=0}(a) and \ref{fig:stat
  IET Hawkes p=0}(b) are also shown in the figure. The introduction of
$p>0$ does not decorrelate the CV and IET correlation.
To explain the behavior of the IET correlation in the present data,
we need different models.
It seems that the IET correlation has not been discussed in the context of social interaction data,
with a notable exception \cite{Karsai2012SciRep}. We are interested in the capabilities of 
alternative models \cite{HanZhouWang2008NJP,Malmgren2008PNAS,Malmgren2009Science} in reproducing 
the IET correlation in the data.

In the present study, we used the exponential memory kernel because it is
analytically tractable and contains only three parameters.
The original Hawkes process with other memory kernels
has also been applied to data
\cite{Ogata1999PAG,Crane2008PNAS}. The ML method is available also for this case 
\cite{Ogata1999PAG}.
Nevertheless, we suspect that self-excitation inherent in the Hawkes process
induces both high CV and positive IET correlation for a variety of memory kernels.
Therefore, the use of different memory kernel
may not improve the fit of the Hawkes process to our data in
terms of the IET correlation.

Two-state models \cite{Malmgren2008PNAS,Malmgren2009Science,Karsai2012SciRep}, in which events are produced at high
and low rates in the
excited and normal states, respectively, are also
self-exciting.
These models may be more realistic for social data than the standard Hawkes process
used in this work in the sense that humans may not 
distinguish many different levels of self-excitation as is assumed in
the Hawkes process. On the other hand,
the Hawkes process with the exponential memory kernel is simpler than these models
such that the ML methods are available and
the parameters have simple physical meanings.
Although the model by Malmgren and colleagues allows for the ML method
\cite{Malmgren2008PNAS}, the method is quite complicated and contains
many parameters.
It may be desirable to develop two-state models that are simple and allow for statistical methods. Alternatively, it may be desirable to modify the Hawkes
process to account for the behavior of the IET correlation in the real data.

We lack methods to compare the goodness of fit of
different models, except that it is straightforward to test the
validity of a model against the Poisson process (but see \cite{Malmgren2008PNAS}). We need develop goodness of fit tests
to compare the performance of models proposed in different papers.

\section*{Appendix 1: IET Distribution of the Hawkes Process}
\addcontentsline{toc}{section}{Appendix}

In this section, we explain the derivation of the IET distribution
of the Hawkes process shown in \cite{Hawkes1974JAP}.
Also see \cite{Verejones1970JRSSB} for introduction to mathematical treatments of the Hawkes and related processes.

Consider a trigger event at $t=0$ and
the inhomogeneous Poisson process
with rate function $\phi(t)$, i.e., the point process directly induced by the trigger event.
The probability generating functional (PGFL) 
for this inhomogeneous Poisson process, denoted by $H$, is given by
\begin{align}
H\left(z(\cdot)\right) \equiv& E\left(\prod_{i\ge 1} z(t_i)\right)\notag \\
=& \exp\left\{ \int_{0}^{\infty}\left[z(t)-1\right]\phi(t) dt\right\},
\label{eq:H(z)}
\end{align}
where $z(t)$ is a carrying function, and $t_i$ is the time of the $i$th event. We define
$t_0= 0$.

The events at $t=t_i$ may induce further events. On the basis of
\EQ\eqref{eq:H(z)}, the PGFL
for the inhomogeneous Poisson process
including all the descendant events induced by a trigger event at $t=0$, denoted by $F$,
is given through the following recursive relation:
\begin{equation}
F\left(z(\cdot)\right) = z(0)\exp\left\{\int_0^{\infty}\left[
F\left(z_t(\cdot)\right)-1\right]\phi(t)dt \right\},
\label{eq:F(z)}
\end{equation} 
where $z_t(t^{\prime})\equiv z(t^{\prime}+t)$ is the time translation.
On the right-hand side of \EQ\eqref{eq:F(z)},
$z(0)$ accounts for the trigger event at $t=0$, and $F\left(z_t(\cdot)\right)$ accounts for the fact that an event triggered at time $t$ initiates an inhomogeneous Poisson process with rate $\phi(t)$ on top of the other inhomogeneous Poisson processes going on.

We obtain the PGFL for the entire Hawkes process, denoted by $G$, by
combining \EQ\eqref{eq:F(z)} and the PDFL of the homogeneous Poisson process with rate $\nu$
as follows:
\begin{equation}
G\left(z(\cdot)\right) = \exp\left\{\int_{-\infty}^{\infty}\nu
\left[F\left(z_t(\cdot)\right)-1 \right]dt\right\}.
\label{eq:G(z)}
\end{equation}

We set $z(t)=\tilde{z}$ for $t_{\rm s}\le t\le t_{\rm s}+\Delta$ and $z(t)=1$
otherwise. Then, $\pi(t_{\rm s}, \Delta, \tilde{z})\equiv F(z(\cdot))$ is the
probability generating function (PGF) for the number of events in $[t_{\rm s}, t_{\rm s}+\Delta]$, with the carrying variable $\tilde{z}$, and 
\begin{equation}
\pi(t_{\rm s}-t, \Delta, \tilde{z}) = F(z_t(\cdot))
\label{eq:F and pi}
\end{equation}
is the PGF for the number of events in $[t_{\rm s}-t, t_{\rm s}-t+\Delta]$.
Equation~\eqref{eq:F(z)} is reduced to
\begin{equation}
\pi(t_{\rm s}, \Delta, \tilde{z})=\begin{cases}
\exp\left\{\int_0^{t_{\rm s}+\Delta}\left[\pi(t_{\rm s}-t, \Delta, \tilde{z})-1
  \right]\phi(t) dt \right\},& t_{\rm s}>0,\\
\tilde{z}\exp\left\{\int_0^{t_{\rm s}+\Delta}\left[\pi(t_{\rm s}-t, \Delta, \tilde{z})-1
  \right]\phi(t) dt \right\},& -\Delta\le t_{\rm s}\le 0,\\
1, & t_{\rm s}<-\Delta.
\end{cases}
\label{eq:pi}
\end{equation}
By setting $t_{\rm s}=0$ and combining \EQS\eqref{eq:G(z)} and \eqref{eq:F and pi},
we obtain the PGF for the number of events in $[0,\Delta]$
as
\begin{equation}
Q_{\Delta}(\tilde{z})\equiv G(z(\cdot)) = 
\exp \left\{ \int_{-\infty}^{\Delta}\nu\left[\pi(-t, \Delta, \tilde{z})-1\right]dt \right\}.
\label{eq:Q(tilde(z))}
\end{equation}

In particular,
\begin{equation}
\tilde{\pi}(t_{\rm s}, \Delta)\equiv \pi(t_{\rm s}, \Delta,0)
\end{equation}
is the probability that there is no event in $[t_{\rm s},
t_{\rm s}+\Delta]$ for a cluster of events originating at $t=0$. Using
\EQ\eqref{eq:pi}, we obtain
\begin{equation}
\tilde{\pi}(t_{\rm s}, \Delta) =
\begin{cases}
\exp\left\{ \int_0^{t_{\rm s}+\Delta}\left[\tilde{\pi}(t_{\rm s}-t,\Delta)-1
\right]\phi(t)dt\right\}, & t_{\rm s}>0,\\
0, & -\Delta\le t_{\rm s}\le 0,\\
1, & t_{\rm s}<-\Delta.
\label{eq:tilde(pi)}
\end{cases}
\end{equation}

By setting $\tilde{z}=0$ in \EQ\eqref{eq:Q(tilde(z))} and using \EQ\eqref{eq:tilde(pi)},
we obtain the survivor function of the
forward recurrence time, i.e., time to the next event
from arbitrary $t$, as follows:
\begin{equation}
Q_{\Delta}(0) = {\rm Pr}(\mbox{forward recurrence time }>\Delta) = \exp \left\{-\nu\Delta -\nu
\int_0^{\infty}\left[ 1-\tilde{\pi}(t, \Delta) \right]dt \right\},
\label{eq:forward recurrence time}
\end{equation}
where ${\rm Pr}$ denotes probability.
$Q_{\Delta}(0)$ is the probability that the Hawkes process does not have
any event in $[0, \Delta]$.

Finally, the distribution of the interevent time $\tau$ is given in the form
of survivor function as
\begin{equation}
{\rm Pr}(\tau>t) = -\frac{dQ_{\Delta}(0)(t)}{dt}\bigg/ \overline{\lambda},
\label{eq:tau survival}
\end{equation}
where the stationary event rate $\overline{\lambda}$ is given by \EQ\eqref{eq:overline(lambda)}.

In the numerical simulations, we adopted the Simpson's rule for calculating integrals
in \EQS\eqref{eq:tilde(pi)} and \eqref{eq:forward recurrence time}, and solved 
\EQ\eqref{eq:tilde(pi)} by iteration.

We remark that integration of \EQ\eqref{eq:tau survival} by part leads
to
\begin{equation}
\left<\tau\right> = \frac{1}{\overline{\lambda}}
\label{eq:<tau> analytical}
\end{equation}
and
\begin{equation}
\left<\tau^2\right>=\frac{2}{\overline{\lambda}}\int_0^{\infty}Q_{\Delta}(0)(t)dt.
\label{eq:<tau^2> analytical}
\end{equation}
Equations~\eqref{eq:<tau> analytical} and \eqref{eq:<tau^2>
  analytical} can serve to calculate the CV. However, we did not use them and
obtained the CV by direct numerical simulations because calculating
the CV via \EQS~\eqref{eq:<tau> analytical} and \eqref{eq:<tau^2> analytical} is time consuming.

\section*{Appendix 2: ML Method for the Hawkes Process}\label{sec:ML method}

In this section, we explain a slightly modified version of 
the ML method for the Hawkes process with the exponential memory kernel
originally proposed in
\cite{Ozaki1979AISM}.

We let the event times be
$0\le t_1 \le t_2 \le \cdots \le t_N$. Different from the usual assumption
of the continuous-time point process, we allow multiple events to
occur at the same time (i.e., $t_i=t_{i+1}$). Such simultaneous events actually
occur in our data because of the finite time resolution of one minute.
Simultaneous events do not prevent the application of the ML method explained in the following.

For the exponential memory kernel
given by \EQ\eqref{eq:phi def exponential},
the event rate at time $t$ is given by
\begin{equation}
\lambda (t) = \nu + \alpha \sum_{j=1}^{j_{\max}(t)}e^{-\beta (t-t_j)},
\label{eq:Lambda}
\end{equation}
where $j_{\max}(t)$ is the index of the last event before time $t$.

The likelihood of the event sequence during time period
$[0,t_N]$, denoted by $L(t_1, \ldots, t_N)$, is given by
%
%
\begin{equation}
L(t_1, \ldots, t_N) = \exp\left( -\int_0^{t_N}\lambda(t)dt
\right)\prod_{i=1}^N \lambda (t_i).
\label{eq:L}
\end{equation}
By substituting
\EQ\eqref{eq:Lambda} in \EQ\eqref{eq:L}, we obtain the log likelihood
for the original Hawkes process as follows \cite{Ozaki1979AISM}:
\begin{equation}
\log L(t_1, \ldots, t_N) = -\nu t_N + \sum_{i=1}^N\frac{\alpha}{\beta}
\left(e^{-\beta (t_N-t_i)}-1 \right) + \sum_{i=1}^N \log (\nu + \alpha A(i)),
\label{eq:log L}
\end{equation}
where
\begin{equation}
A(i) = \sum_{1\le j<i \le N}e^{-\beta (t_i-t_j)}.
\end{equation}

Exactly speaking, the point process for an individual for one workday
begins when the individual has arrived in the office.
Because we do not know when the point process begins, we
assume that the first event of each day is a trigger event. 
In other words, we set $t_1=0$ and modify
\EQ\eqref{eq:log L} as
\begin{equation}
\log L(t_1, \ldots, t_N) = -\nu t_N + \sum_{i=1}^N\frac{\alpha}{\beta}
\left(e^{-\beta (t_N-t_i)}-1 \right) + \sum_{i=2}^N \log (\nu + \alpha A(i)).
\label{eq:log L modified}
\end{equation}

For each individual, we
use the days that have at least 40 events. We index such a valid day
as $d=1, 2, \ldots, d_{\max}$. We denote the event times of
valid day $d$ by
$0= t^d_1\le \ldots \le t^d_{N_d}$, where $N_d$ is the number of events in
valid day $d$. The log likelihood of the entire sequence is given by
the summation of the log likelihood over all the valid days.

The partial derivatives of the log likelihood
with respect to $\alpha$, $\beta$, and $\nu$ are
originally derived in \cite{Ozaki1979AISM}.
In the present case, they read
\begin{align}
\frac{\partial \log L}{\partial\alpha} =&
\sum_{d=1}^{d_{\max}} \left\{
\sum_{i=1}^{N_d} \frac{1}{\beta}\left(e^{-\beta (t_{N_d}^d-t_i^d)}-1
\right)
+ \sum_{i=2}^{N_d}\frac{A_d(i)}{\nu+\alpha A_d(i)}
\right\},\label{eq:dL/dalpha}\\
\frac{\partial \log L}{\partial\beta} =&
\sum_{d=1}^{d_{\max}} \left\{
-\alpha\sum_{i=1}^{N_d}\left[ \frac{1}{\beta}(t_{N_d}^d-t_i^d)
e^{-\beta(t_{N_d}^d-t_i^d)}
+ \frac{1}{\beta^2}\left(e^{-\beta(t_{N_d}^d-t_i^d)}-1\right)
\right]-\right.\notag\\
& \left. \sum_{i=2}^{N_d}\frac{\alpha B_d(i)}{\nu+\alpha A_d(i)}
\right\},\label{eq:dL/dbeta}\\
\frac{\partial \log L}{\partial\nu} =& 
\sum_{d=1}^{d_{\max}} \left\{ -t_{N_d}^d +
  \sum_{i=2}^{N_d}\frac{1}{\mu+\alpha A_d(i)}
\right\},\label{eq:dL/dnu}
\end{align}  
where
\begin{equation}
A_d(i) = \sum_{1\le j < i \le N_d} e^{-\beta (t_i^d-t_j^d)}
\end{equation}
and
\begin{equation}
B_d(i) = \sum_{1\le j < i \le N_d} (t_i^d-t_j^d)e^{-\beta (t_i^d-t_j^d)}.
\end{equation}
We obtain the ML estimates by
setting the left-hand sides of \EQS\eqref{eq:dL/dalpha},
\eqref{eq:dL/dbeta}, and \eqref{eq:dL/dnu} to 0.

We carried out the gradient descent method to estimate $\alpha$, $\beta$, and
$\nu$ for each individual.
We repeat the substitution
\begin{align}
\alpha \gets& \alpha + \delta \frac{\partial \log L}{\partial \alpha},\\
\beta \gets& \beta + \delta \frac{\partial \log L}{\partial \beta},\\
\nu \gets& \nu + \delta \frac{\partial \log L}{\partial \nu},
\end{align}
where we set $\delta=10^{-2}$. For one individual in $D_2$, the ML method
does not converge with $\delta=10^{-2}$. Because it converges
with $\delta=10^{-3}$, we used this value for this particular individual.

Because the likelihood may have multiple local maxima,
we started the
gradient descent method with two different
initial conditions, i.e., $(\alpha, \beta, \nu)=(0.6, 1.2, 0.6)$ and
$(12, 24, 12)$ [hr$^{-1}$]. We found that the final results corresponding to
the two initial conditions were identical for each individual.

For the ML method, the Hessian of the log likelihood
can be explicitly given and used in combination with the Newton method 
\cite{Ozaki1979AISM}. However, we found that 
the Newton method does not converge 
for many individuals compared to the 
simple gradient descent described above. Therefore, we did not use
the Newton method.

Because $\alpha, \beta, \nu\ge 0$ and $\alpha<\beta$ 
are needed for the Hawkes process to be well defined,
we forced the parameter values to satisfy these conditions. In each
update step, if the updated $\alpha$ becomes less than $10^{-6}$, we set
$\alpha=10^{-6}$. Similarly, if $\alpha<\beta$ is violated, we set
$\beta=\alpha+10^{-6}$. If $\nu<10^{-6}$, we set
$\nu=10^{-6}$.

The temporal resolution of our data is a minute. We set the unit time
for the ML method to an hour such that our data has a resolution of
$1/60$ on this timescale. The data would be too discrete for the ML method to bear accurate results
if we set the unit time for the ML method to a minute.
We verified that the results little changed when we made the time
unit larger than one hour.



%
%

\begin{acknowledgement}
N. M. acknowledges the support provided through Grants-in-Aid for Scientific Research (No. 23681033, and Innovative Areas ``Systems Molecular Ethology''(No. 20115009)) from MEXT, Japan. 
T. T. acknowledges the support provided through
Grants-in-Aid for Scientific Research (No. 10J06281)
from JSPS, Japan.
\end{acknowledgement}

\newpage
\clearpage

\clearpage

\clearpage

\clearpage

\clearpage

\clearpage

\clearpage

\clearpage

\end{document}